\documentclass[12pt, reqno]{article}
\usepackage{epsfig}
\usepackage{amssymb}
\usepackage{amsmath}
\usepackage{mathrsfs}
\usepackage{todonotes}
\usepackage{slashed}
\usepackage{hyperref}

\newcommand{\half}{\frac{1}{2}}

\newcommand{\M}{{\cal{M}}}
\newcommand{\e}{\epsilon}

\newcommand{\be}{\begin{eqnarray}}
\newcommand{\ee}{\end{eqnarray}}

\def\cs{{\cal S}}
\def\pl{\lambda}
\begin{document}

\baselineskip=18pt

\setcounter{footnote}{0}
\setcounter{figure}{0}
\setcounter{table}{0}

\begin{titlepage}
\unitlength = 1mm
\ \\

\today
\begin{center}

{ \LARGE {\textsc{Evidence for a New Soft Graviton Theorem}}}

\vspace{0.8cm}
Freddy Cachazo$^*$ and Andrew Strominger$^\dagger$

\vspace{1cm}
\begin{abstract}The single-soft-graviton limit of any quantum gravity scattering amplitude is given at leading order by the  universal Weinberg pole formula. Gauge invariance of the formula follows from global energy-momentum conservation.  In this paper evidence is given for a conjectured universal formula for the finite subleading term in the expansion about the soft limit, whose gauge invariance follows from global  angular momentum conservation. The conjecture is non-trivially verified for all tree-level graviton scattering amplitudes using a BCFW recursion relation. One hopes to understand this infinity of new soft relations as a Ward identity for a new superrotation Virasoro symmetry of the quantum gravity $\cs$-matrix.      \end{abstract}
\vspace{0.5cm}

\vspace{1.0cm}

\end{center}
\vspace{6.0cm}
$^*${\it Perimeter Institute for Theoretical Physics, Waterloo, ON, Canada} \\
$^\dagger$ {\it Center for the Fundamental Laws of Nature, Harvard University,
Cambridge MA, USA}

\end{titlepage}

\pagestyle{empty}
\pagestyle{plain}

\def\ip{${\cal I}^+$}
\def\p{\partial}
\def\cs{{\cal S}}

\def\co{{\cal O}}

\pagenumbering{arabic}

\tableofcontents

\section{Introduction}

A generic (n+1)-particle on-shell scattering amplitude including  an external graviton with momentum $q$ may be denoted
\be {\cal M}_{n+1} = \M_{n+1}(k_1,k_2,...k_n,q) .\,
\ee
Weinberg's theorem \cite{steve} states that near the $q\to 0$ soft limit
\be \label{wf}\M_{n+1}(k_1,k_2,...k_n,q) = S^{(0)}\M_{n}(k_1,k_2,...k_n)+\co(q^0) ,\ee
where the leading soft factor is a pole\footnote{ Here the soft graviton polarization tensor obeys $E_{\mu\nu} q^\nu=0$ and $8\pi G=1$.}
\be S^{(0)}\equiv \sum_{a=1}^n {E_{\mu\nu}k_a^\mu k_a^\nu\over q\cdot k_a}.\ee
 Gauge invariance requires that
the pole term  vanish for $ \delta_\Lambda E_{\mu\nu} =\Lambda_\mu q_\nu+\Lambda_\mu q_\nu$ with $\Lambda\cdot q=0$.
Indeed one finds
\be \delta_\Lambda S^{(0)}=2\Lambda_\mu \sum_{a=1}^nk_a^\mu =0\ee
as a consequence of  global-energy momentum conservation.

Recently it was shown \cite{bms1,bms2} that the universal formula (\ref{wf}) in fact follows from a symmetry principle. Gravitational scattering -both classical and quantum -has infinitely many symmetries which form a certain ``anti-diagonal" subgroup\footnote{These  antidiagonal elements involve an antipodal identifications of parameters on the conformal spheres at past and future null infinity.}  of the product group of BMS supertranslations \cite{bms} acting on past and future null infinity.
 (\ref{wf}) is then the Ward identity associated to this infinite-dimensional anti-diagonal symmetry.\footnote{ The logic may be turned around: soft theorems typically imply infinitely many symmetries of the  $\cs$-matrix, and anti-diagonal BMS symmetry can be derived from Weinberg's theorem.}

BMS transformations include, in addition to the supertranslations, $SL(2,\mathbb{C})$ Lorentz transformations that act conformally on the sphere at past or future  null infinity.  It has been conjectured in \cite{bt,Banks:2003vp} that the BMS transformations  should be infinitely extended to include all Virasoro transformations, $i.e.$ superrotations, on the conformal sphere.
In \cite{asc} it was further conjectured that the anti-diagonal subgroup of the past and future Virasoro symmetries are a symmetry of the
$\cs$-matrix, and that there is an associated new soft theorem
\be \label{sf}{\cal M}_{n+1}(k_1,k_2,...k_n,q) = S^{(0)}{\cal M}_{n}(k_1,k_2,...k_n)+S^{(1)}{\cal M}_{n}(k_1,k_2,...k_n)+\co(q^1) ,\ee
where the subleading soft factor is
\be\label{sls} S^{(1)}\equiv -i\sum_{a=1}^n {E_{\mu \nu}k_{a}^{\mu} (q_\rho J_{a}^{\rho\nu})\over q \cdot k_a}, \ee
with $J_a$ the total angular momentum of the $a$th particle.\footnote{In a momentum eigenbasis  this acts as a differential operator $J_{a\mu\nu}\sim k_{a[\mu}{\p \over \p k_{a}^{~\nu]} }$+(helicity terms) on $M_n$. The action of these differential operators on the on-shell amplitudes is spelled out below.} 
A number of considerations led to this conjecture: here we note only that it is gauge invariant
\be  \delta_\Lambda S^{(1)}=-i\Lambda_\mu q_\nu \sum_{a=1}^nJ_a^{\mu\nu} =0 \ee
as a consequence of global angular momentum conservation. Subleading terms in soft expansions have a long history (see e.g. \cite{Low:1958sn,Burnett:1967km, White:2011yy}); some of the first results on graviton scattering off scalars go back to work by Gross and Jackiw \cite{Gross:1968in,Jackiw:1968zza}.

There are several approaches to prove or disprove the conjecture (\ref{sf}). One would be a diagrammatic approach a la
Weinberg, in which one works to one higher order around the soft limit. This is subtle because there are many sources of subleading corrections. Another follows the path of \cite{bms1,bms2} to Weinberg's theorem in explicitly constructing the generators of the conjectured superrotation Virasoro symmetries and rewriting the invariance of the $\cs$-matrix in the form of a soft theorem. This is substantially more complicated than the leading supertranslation case worked out in \cite{bms1,bms2}, but progress in this direction appears promising \cite{klps} and indeed motivated the conjecture (\ref{sf}).

In this paper we employ a third, orthogonal, approach. We will not try to prove (\ref{sf}). Rather we will simply test if it is true for tree-level gravitational scattering amplitudes, using the BCFW construction \cite{Britto:2004ap,bcfwa}. The conjecture (\ref{sf}) is found to pass this tree-level test in a highly nontrivial manner. The test is possible due to the deep progress in understanding these amplitudes and their properties which has evolved in recent years, for a recent review see \cite{Elvang:2013cua}. This progress in turn may have been possible because of the new symmetries pointed to herein.

We employ the spinor helicity formalism which allows us to define and study a certain holomorphic soft limit.   This limit  allows us to see deeper into the expansion around the soft limit and discover a surprising universality in a sub-subleading term. We find by direct computation that all graviton tree amplitudes in fact obey a soft identity even stronger than (\ref{sf}) of the form
\be \label{nextsf}{\cal M}_{n+1}(k_1,k_2,...k_n,q) = \left(S^{(0)}+S^{(1)}+S^{(2)}\right){\cal M}_{n}(k_1,k_2,...k_n)+\co(q^2) ,
\ee
where
\be\label{nexttospin}
S^{(2)} \equiv -\half \sum_{a=1}^n \frac{E_{\mu\nu}(q_\rho J_a^{\rho\mu})(q_\sigma J_a^{\sigma\nu})}{q\cdot k_a}.
\ee
It is easy to check that $S^{(2)}$ is gauge invariant as required for the simple reason that $J_a^{\mu\nu}$ is antisymmetric and not as a consequence of any conservation law. While this result is highly suggestive, unlike $S^{(1)}$, at this point we have found no argument that $S^{(2)}$ is universal
beyond tree-level gravity. We leave the origin of this surprising relation to future investigations.

Going back to the original conjecture (\ref{sf}), several caveats are in order. First, in taking the soft limit of one external graviton, momenta of some of the hard particles must be simultaneously deformed in order to conserve momentum. Hence there are many expansions about the soft limit corresponding to the many possible deformations of the hard particle momenta. We prove (\ref{sf}) applies universally to a very large class of such soft limit expansions which are naturally defined in the spinor helicity formalism. We have not shown that it holds for
every conceivable definition of the soft limit expansion.

Second, since our tests here are purely classical, they leave open the possibility that the conjecture (\ref{sf}) or its extension (\ref{nextsf}) is either invalid or in need of modification at the quantum level.

We hope that future work can address both of these issues.

This paper is organized as follows.  Section 2.1 contains our conventions and some spinor helicity formulae. In section 2.2 we precisely state our claim, emphasizing that the identity (\ref{nextsf}) involves the full physical amplitudes which contain distributional momentum-conserving delta functions. The claim is proven in 2.3 using a factorized form of the amplitudes which played a key role in \cite{ack}.  Section 3 concerns the unphysical, non-distributional ``stripped" amplitudes, which are sometimes easier to work with and commonly considered in the literature. These obey a slightly modified form of
the identity (\ref{nextsf}) which is presented in detail. Section 4 demonstrates the validity of (\ref{nextsf}) in detail for some low-point amplitudes. Appendix A demonstrates the soft regularity of certain terms in the factorized amplitude sum,
Appendix B gives details of a Taylor expansion and Appendix C reviews Hodges' formula for MHV amplitudes.

\section{All tree-level graviton scattering amplitudes}

\subsection{Preliminaries}

Our derivation employs the spinor helicity formalism \cite{Elvang:2013cua}.  This subsection contains  our conventions and a few essential spinor helicity formulae.

Consider any on shell tree level amplitude of $n+1$ gravitons. We will single out one of them which will eventually become soft and label it by $s$ with momentum vector $q$. The remaining $n$ gravitons will be labeled using the standard $\{ 1,2,\ldots ,n\}$ notation.
The kinematic data is given by a pair of spinors for each particle $\lambda_{a,\alpha},\tilde\lambda_{a,\dot\alpha}$. The momentum vector is given as a bispinor
\begin{eqnarray} k_{a,\alpha\dot\alpha} &=& \lambda_{a,\alpha}\tilde\lambda_{a,\dot\alpha},~~~a\in \{1,2\ldots ,n\}\cr q_{\alpha\dot\alpha }&=&\lambda_{s,\alpha}\tilde\lambda_{s,\dot\alpha}.\end{eqnarray} The only extra piece of information needed is the helicity of the particle which we take to be $h=\pm 2$. Momentum conservation implies
\be\label{xnbp}
\sum_{a=1}^n\lambda_{a,\alpha} \tilde\lambda_{a,\dot\alpha}+\lambda_{s,\alpha}\tilde\lambda_{s,\dot\alpha}=0 .
\ee
The amplitude is a function of the form
\be \label{xds}
\M_{n+1} = M_{n+1}(\{\lambda_1,\tilde\lambda_1,h_1\},\ldots ,\{\lambda_n,\tilde\lambda_n,h_n\},\{\lambda_s,\tilde\lambda_s,h_s\}) \delta^4( \sum_{a=1}^n\lambda_{a,\alpha} \tilde\lambda_{a,\dot\alpha}+\lambda_{s,\alpha}\tilde\lambda_{s,\dot\alpha}),
\ee
where $M_{n+1}$ is a  `` stripped amplitude" without a momentum-conserving delta function.
Here $h_i$ denotes the helicity of the $i^{\rm th}$ graviton and from now on we take $h_s=+2$. In the rest of the paper we will suppress the helicity dependence unless needed.

Weinberg's soft factor
\be
S^{(0)}\equiv \sum_{a=1}^n {E_{\mu\nu}k_a^\mu k_a^\nu\over q\cdot k_a}
\ee
can easily be rewritten in spinor helicity form. The first step is to note that for a particle with positive helicity (see e.g. \cite{Witten:2003nn})
\be
E_{\alpha\dot\alpha \beta \dot\beta}= \left(\frac{\lambda_{x,\alpha}\tilde\lambda_{s,\dot\alpha}}{\langle x,s \rangle}\right)\left(\frac{\lambda_{y,\beta}\tilde\lambda_{s,\dot\beta}}{\langle y,s \rangle}\right) + (x\leftrightarrow y).
\ee
Here $\lambda_x$ and $\lambda_y$ are two arbitrary chosen reference spinors, often judiciously chosen to simplify computations. The freedom in this choice is equivalent to the freedom in gauge choice for the polarization tensor. We have also adopted the standard notation
\be [s,a]\equiv \tilde \lambda_{s,\dot\alpha}\tilde\lambda_{a}^{~\dot\alpha},~~~  \langle s,a\rangle \equiv \lambda_{s,\alpha}\lambda_{a}^{~\alpha} .\ee
The inner product of two null vectors is given, e.g., $q$ and $k_a$ by
\be
2q\cdot k_a = \langle s,a\rangle [s,a].
\ee
Weinberg's soft theorem for $q$ soft becomes, in spinor helicity form
\begin{eqnarray}\label{weinberg}\M_{n+1} &\to& S^{(0)}  \M_{n}(\{\lambda_1,\tilde\lambda_1\},\ldots ,\{\lambda_n,\tilde\lambda_n\}),\cr
~~~~~S^{(0)}&=&\sum_{a=1}^n \frac{[s,a]}{\langle s,a\rangle}\frac{\langle x,a\rangle\langle y,a\rangle}{\langle x,s\rangle\langle y,s\rangle}.
\end{eqnarray}
Here $\M_n$ is a fully physical amplitude evaluated on the $n$-particle data obtained by taking $q$ identically equal to zero. The independence of expression on the choice of reference spinors $\lambda_x$ and $\lambda_y$ is a consequence of energy-momentum conservation of the $n$-particle data. The subleading operator appearing in the conjecture (\ref{sls}) becomes,  in spinor notation
\be \label{sone}
S^{(1)} = \frac{1}{2}\sum_{a=1}^{n}\frac{[s,a]}{\langle s,a\rangle}\left(\frac{\langle x,a\rangle}{\langle x,s\rangle}+\frac{\langle y,a\rangle}{\langle y,s\rangle}\right)\tilde\lambda_s^{\dot\alpha}\frac{\partial}{\partial\tilde\lambda_a^{\dot\alpha}}.
\ee
Here we have used\footnote{Here the normalization is $\sigma^{\alpha \dot\alpha}_\mu \sigma_{\nu \alpha \dot\alpha}=2\eta_{\mu\nu}$.}
\be
J_{\mu\nu}\sigma^\mu_{\alpha \dot\alpha}\sigma^\nu_{\beta\dot\beta}= -2J_{\alpha\beta}\epsilon_{\dot\alpha \dot\beta} -2\epsilon_{\alpha \beta}{\tilde J}_{\dot\alpha \dot\beta}
\ee
with
$$J_{\alpha\beta} =\frac{i}{2} \left(\lambda_\alpha\frac{\partial}{\partial\lambda^\beta}+ \lambda_\beta\frac{\partial}{\partial\lambda^\alpha}\right), \qquad {\tilde J}_{\dot\alpha \dot\beta} = \frac{i}{2}\left( \tilde\lambda_{\dot\alpha}\frac{\partial}{\partial\tilde\lambda^{\dot\beta}}+ \tilde\lambda_{\dot\beta}\frac{\partial}{\partial\tilde\lambda^{\dot\alpha}}\right).$$
Finally, using this formula,  the second order soft operator in (\ref{nexttospin}) becomes
\be \label{stwo}
S^{(2)} =  \frac{1}{2}\sum_{a=1}^{n}\frac{[s,a]}{\langle s,a\rangle}\tilde\lambda_s^{\dot\alpha}\tilde\lambda_s^{\dot\beta}\frac{\partial^2}{\partial \tilde\lambda_a^{\dot\alpha}\partial \tilde\lambda_a^{\dot\beta}}.
\ee

\subsection{Claim}

We define the soft limit expansion of $M_{n+1}$ by introducing a small expansion parameter $\epsilon$ which multiplies the momentum of the soft graviton:%
\be
q_{\alpha,\dot\alpha} \to \epsilon q_{\alpha,\dot\alpha} = \epsilon\lambda_{s,\alpha}\tilde\lambda_{s,\dot\alpha}.
\ee
We must further specify  how $\lambda_s$ and $\tilde\lambda_s$ are taken to zero. A natural  choice is
\be\label{realdeform}
\lambda_{s}\to \sqrt{\epsilon}\lambda_{s},\quad \tilde\lambda_{s} \to \sqrt{\epsilon}\tilde\lambda_{s}.
\ee
The precise form of our claim in spinor helicity form is then
\be\label{minclaim}
{\cal M}_{n+1}(...,\{\sqrt{\epsilon} \lambda_s,\sqrt{\epsilon}\tilde\lambda_s\}) = \left({1\over \e}S^{(0)} + S^{(1)}+ \e S^{(2)}\right){\cal M}_n+\co (\epsilon^2).
\ee
In this expression, the factors of $\e$ are explicitly displayed and $S^{(i)}$ are understood to be given exactly by (\ref{weinberg}), (\ref{sone}) and (\ref{stwo})
without any rescaling of $\lambda_s, \tilde \lambda_s$ appearing therein.
The left hand side of this equation involves the (unstripped distributional) $(n+1)$-particle  amplitude
\be
{\cal M}_{n+1}(...,\{\sqrt{\epsilon}\lambda_s,\sqrt{\epsilon}\tilde\lambda_s\}) = M_{n+1}(\{\lambda_1,\tilde\lambda_1\},\ldots , \{\lambda_n,\tilde\lambda_n\},\{\sqrt{\epsilon}\lambda_s,\sqrt{\epsilon}\tilde\lambda_s\})
\delta^4(\sum_{a=1}^nk_n^\mu+\epsilon q^\mu)
\ee
while  right hand side involves the $n$-particle amplitude
\be
{\cal M}_n = M_{n}(\{\lambda_1,\tilde\lambda_1\},\ldots , \{\lambda_n,\tilde\lambda_n\})
\delta^4(\sum_{a=1}^nk_n^\mu).
\ee

The deformation (\ref{realdeform}) has the property  that if the original momentum of the soft particle is real, i.e., $\tilde\lambda_s = \pm (\lambda_s)^*$ then it stays real for all real values of $\epsilon$. This condition might be important in analyzing loop level amplitudes but since we are only interested in tree-level amplitudes, which are rational functions, this restriction is not necessary. In fact we will find that it is technically more convenient to study a holomorphic soft limit as in  \cite{ack}. To define this limit recall that any physical amplitude containing a particle of helicity $h$ transforms under the little group as
\be
\M(\{t\pl,t^{-1}\tilde\pl, h \}) = t^{-2h}\M(\{\pl ,\tilde\pl, h\}).
\ee
Applying this to particle $s$ in $M_{n+1}(\epsilon )$ with $t=\epsilon^{-1/2}$ one finds
\be\label{rell}
\M_{n+1}(\{\sqrt{\epsilon}\pl_s,\sqrt{\epsilon}\tilde\pl_s,+2\}) = \epsilon^{2}\M(\{\epsilon \pl_s,\tilde\pl_s,+2\}).
\ee
The holomorphic soft limit is the usual soft limit combined with a little group transformation in such a way that only the holomorphic spinor $\pl_s(\epsilon )\to 0$ while the anti-holomorphic one, $\tilde\pl_s$, stays finite and generic. One then finds that the claim (\ref{minclaim}) is equivalent to
\be \label{mainclaim}
{\cal M}_{n+1}(...,\{ \epsilon \lambda_s, \tilde\lambda_s\}) = \left({1 \over \e^3}S^{(0)} + {1 \over \e^2}S^{(1)}+ {1 \over \e}S^{(2)}\right){\cal M}_n+\co (\epsilon^0 ).
\ee
Note that the undetermined part is now finite, while the universal parts have poles up to cubic order.

\subsection{Proof}

The proof of the claim (\ref{mainclaim}) is achieved in several steps. First we find a convenient factorized representation of the general stripped $n+1$ graviton amplitude $M_{n+1}$, in which only certain special terms can develop poles in $\epsilon$. The second is to apply the deformation (\ref{realdeform}) to this representation to get $\M_{n+1}(\epsilon )$.  Equation~\ref{mainclaim} is then recovered via a Taylor expansion of the special terms. Our construction uses the same BCFW \cite{Britto:2004ap,bcfwa} analysis employed in the study of a single soft emission in \cite{ack}.

\subsubsection{Factorized representation of $M_{n+1}$}

Consider a  BCFW-type expansion of a general stripped amplitude $M_{n+1}$ based on the deformation
\be
\lambda_s(z) = \lambda_s + z\lambda_n, \quad \tilde\lambda_n(z) = \tilde\lambda_n - z \tilde\lambda_s.
\ee
where the $\lambda$s obey (\ref{xnbp}). Under this deformation the amplitude becomes a rational function of $z$ denoted $M_{n+1}(z)$. Clearly, the original amplitude is recovered when $z=0$: i.e.
\be\label{bcfw}
M_{n+1} = \frac{1}{2\pi i}\oint_{|z|=\varepsilon}\frac{dz}{z}M_{n+1}(z).
\ee
There are several proofs in the literature \cite{Benincasa:2007qj,ArkaniHamed:2008yf} that $M_{n+1}(z) \to 1/z$ as $z\to \infty$. Therefore the only poles of the integrand are at the solutions of
\be
(k_s(z)+k_{a_1}+k_{a_2}+\cdots + k_{a_m})^2 = 0
\ee
for any non-empty subset $\{a_1,a_2,\ldots, a_m\}$ of $\{1,2,\ldots ,n-1\}$. Note that this does not include the $n^{\rm th}$ graviton.

The residue at each of these poles is determined by unitarity to be the product of two lower point amplitudes \cite{Bedford:2005yy,Cachazo:2005ca}. For a pole at say $z=z^*$, the vector $k_I = k_s(z^*)+k_{a_1}+k_{a_2}+\cdots + k_{a_m}$ is null by construction, and the residue is given by
\be
\sum_{h_I}M_L(s(z^*),a_1,a_2,\ldots ,a_m,\{k_I,h_I\})\frac{1}{P_I^2}M_R(\{-k_I,-h_I\},n(z^*),\overline{a_1,a_2,\ldots ,a_m}) \nonumber
\ee
where $\overline{a_1,a_2,\ldots ,a_m}$ is the complement of $\{a_1,a_2,\ldots, a_m\}$ in $\{1,2,\ldots ,n-1\}$ and $P_I=k_s+k_{a_1}+\cdots + k_{a_m}$. Here $M_L$ and $M_R$ denote the ``Left" and ``Right" amplitudes in the factorization at the pole. The sum is over all possible intermediate states which in this case we take to be gravitons with helicity $h_I=\pm 2$.

Using the residue theorem on (\ref{bcfw}) the amplitude $M_{n+1}$ can be written as a sum over all poles different from $z=0$,
\be\label{fullamp}
M_{n+1}=\sum_{{\cal I},h_I} M_L(s(z^*),a_1,a_2,\ldots ,a_m,\{k_I,h_I\})\frac{1}{P_I^2}M_R(\{-k_I,-h_I\},n(z^*),\overline{a_1,a_2,\ldots ,a_m})
\ee
where the sum is over all non-empty subsets ${\cal I}=\{a_1,a_2,\ldots, a_m\}$ of $\{1,2,\ldots ,n-1\}$.

Let us separate the terms in the sum into two groups. The first group is those where the set ${\cal I}$ consists of a single element and the second group contains all other sets. The reason for this separation is that, as will be proven in Appendix A,  elements of the former becomes singular when $\lambda_s\to 0$ while elements of the latter remains finite.

Let us carefully study the terms in (\ref{fullamp}) of the first kind, i.e., when the set ${\cal I}$ consists of a single element,
\be\label{propa}
\sum_{a=1}^{n-1} \sum_{h_I} M_{L}(s(z^*),a,\{k_I,h_I\})\frac{1}{(k_s+k_a)^2}M_{R}(\{-k_I,-h_I\},n(z^*),\overline{a}).
\ee
In these terms it is easy to compute
\be
z^* = -\frac{\langle a,s\rangle}{\langle a,n \rangle}, \quad \lambda_I = \lambda_a,\quad \tilde\lambda_I = \tilde\lambda_a+\frac{\langle n,s\rangle}{\langle n,a \rangle}\tilde\lambda_s,
\ee
where a potential multiplicative factor in the last two equalities has been set to unity by a little group transformation of
$(\lambda_I, \tilde \lambda_I )$.
Using this it is also possible to show that the  $h_I = h_a$ term vanishes while the $h_I = -h_a$ term gives
\be
M_{L}(\{s(z^*),+2\},a,\{k_I,-h_a\} ) = \frac{[s,a]^2\langle n,a\rangle^2}{\langle n,s\rangle^2}
\ee
regardless of the helicity of the $a^{\rm th}$ graviton. Here we have exhibited the helicity $h_s=+2$ only for the reader's convenience.

Using the fact that the propagator in (\ref{propa}) can be written as $1/\langle s,a \rangle[s,a]$ , (\ref{propa}) may be reduced to
\be
\sum_{a=1}^{n-1} \frac{[s,a]\langle n,a \rangle^2}{\langle s,a \rangle\langle n,s \rangle^2} M_{n}(\{\lambda_1,\tilde\lambda_1,h_1\},\ldots ,\{\lambda_a,\tilde\lambda_a\! + \!\frac{\langle n,s \rangle}{\langle n,a \rangle}\tilde\lambda_s,h_a\},\ldots ,\{\lambda_n,\tilde\lambda_n\! + \!\frac{\langle a,s \rangle}{\langle a,n \rangle}\tilde\lambda_s,h_n\} ).
\ee
Note how the intermediate particle $I$ in $M_R$, with data $\{\lambda_I,\tilde\lambda_I,-h_I\}$, beautifully becomes the $a^{\rm th}$ particle of an $n$-particle amplitude with data $\{\lambda_a,\tilde\lambda_a+\frac{\langle n,s \rangle}{\langle n,a \rangle}\tilde\lambda_s,h_a\}$. Moreover this $n$-particle amplitude is physical in that its $n$-particle data satisfy momentum conservation!

The full amplitude can now be written as
\be\label{repre}
M_{n+1}\!\! =\!\! \sum_{a=1}^{n-1} \frac{[s,a]\langle n,a \rangle^2}{\langle s,a \rangle\langle n,s \rangle^2} M_{n}(\{\lambda_1,\tilde\lambda_1\},\ldots ,\{\lambda_a,\tilde\lambda_a\! + \!\frac{\langle n,s \rangle}{\langle n,a \rangle}\tilde\lambda_s\},\ldots ,\{\lambda_n,\tilde\lambda_n\! + \!\frac{\langle a,s \rangle}{\langle a,n \rangle}\tilde\lambda_s\} )\! + \! {\cal R}
\ee
where ${\cal R}$ stands for terms that are shown in Appendix A to be regular for $\lambda_s\to 0$.

It is interesting that the argument of the $n$-particle amplitude in (\ref{repre}) seems to be obtained by ``dissolving" particle $s$ into particles $a$ and $n$. Similar constructions \cite{BFRSV:2005,BFRSV:2011,Hodges:2011wm,Hodges:2012ym} have appeared in many different applications to the computation of scattering amplitudes of both Yang-Mills and gravity.

\subsubsection{Soft limit expansion}

 In order to obtain expressions for unstripped amplitudes, we dress both sides of (\ref{repre}) by the appropriate momentum-conserving delta function which obeys the identity
\be \label{dfn}
\delta^4(\sum_{b=1}^n\lambda_b\tilde\lambda_b+\lambda_s\tilde\lambda_s)=\delta^4(\sum_{b=1,b\neq a}^{n-1}\lambda_b\tilde\lambda_b+\lambda_a(\tilde\lambda_a+\frac{\langle n,s \rangle}{\langle n,a \rangle}\tilde\lambda_s)+\lambda_n(\tilde\lambda_n+\frac{\langle a,s \rangle}{\langle a,n \rangle}\tilde\lambda_s)).\ee
Using the basic relation (\ref{xds}) between stripped and unstripped amplitudes, and multiplying  the left (right ) hand side
of (\ref{repre}) by the left (right ) hand side of (\ref{dfn}) we immediately obtain
\be\label{repre3}
{\cal M}_{n+1}\!\! =\!\! \sum_{a=1}^{n-1} \frac{[s,a]\langle n,a \rangle^2}{\langle s,a \rangle\langle n,s \rangle^2} {\cal M}_{n}(\{\lambda_1,\tilde\lambda_1\},\ldots ,\{\lambda_a,\tilde\lambda_a+\frac{\langle n,s \rangle}{\langle n,a \rangle}\tilde\lambda_s\},\ldots ,\{\lambda_n,\tilde\lambda_n+\frac{\langle a,s \rangle}{\langle a,n \rangle}\tilde\lambda_s\} ) .
\ee
Here and in the rest of this section we drop $\cal R$ as it does not contribute to the order we are interested in.

The next step is  to apply the deformation $\lambda_s\to \epsilon\pl_s$ to the full $(n+1)$-particle amplitude in the factorized form (\ref{repre3}).
For this  all we have to do is send $\lambda_s$ to $\epsilon\lambda_s$ and Taylor expand ${\cal M}_n$ around $\epsilon =0$ to get
\begin{eqnarray}
&& {\cal M}_{n}(\{\lambda_1,\tilde\lambda_1\},\ldots ,\{\lambda_a,\tilde\lambda_a+\epsilon\frac{\langle n,s \rangle}{\langle n,a \rangle}\tilde\lambda_s\},\ldots ,\{\lambda_n,\tilde\lambda_n+\epsilon\frac{\langle a,s \rangle}{\langle a,n \rangle}\tilde\lambda_s\} )  = \cr
&& \left( 1 + \epsilon \frac{\langle n,s\rangle}{\langle n,a\rangle}\tilde\lambda_s^{\dot\alpha}\frac{\partial}{\partial \tilde\lambda_a^{\dot\alpha}}+ \epsilon \frac{\langle a,s\rangle}{\langle a,n\rangle}\tilde\lambda_s^{\dot\alpha}\frac{\partial}{\partial \tilde\lambda_n^{\dot\alpha}}\right){\cal M}_{n}(\{\lambda_1,\tilde\lambda_1\},\ldots ,\{\lambda_a,\tilde\lambda_a\},\ldots ,\{\lambda_n,\tilde\lambda_n\} )+\co (\epsilon^2).\nonumber
\end{eqnarray}
Plugging this back into the $\epsilon$ deformed version of (\ref{repre}) it is easy to see that the terms with derivatives with respect to $\tilde\lambda_n$ add up to zero.

Putting all of this together gives
\begin{eqnarray}\label{xzd}
&& {\cal M}_{n+1}(\ldots ,\{\epsilon\lambda_s,\tilde\lambda_s\}) = \cr
&& \frac{1}{\epsilon^3}\sum_{a=1}^{n-1} \frac{[s,a]\langle n,a \rangle^2}{\langle s,a \rangle\langle n,s \rangle^2}{\cal M}_{n}(\{\lambda_1,\tilde\lambda_1\},\ldots ,\{\lambda_a,\tilde\lambda_a\},\ldots ,\{\lambda_n,\tilde\lambda_n\} ) \cr && +\frac{1}{\epsilon^2}\sum_{a=1}^{n-1} \frac{[s,a]\langle n,a \rangle}{\langle s,a \rangle\langle n,s \rangle}\tilde\lambda_s^{\dot\alpha}\frac{\partial}{\partial \tilde\lambda_a^{\dot\alpha}}{\cal M}_{n}(\{\lambda_1,\tilde\lambda_1\},\ldots ,\{\lambda_a,\tilde\lambda_a\},\ldots ,\{\lambda_n,\tilde\lambda_n\} )+\co (\epsilon^{-1}).
\end{eqnarray}
The analysis of the order $ 1 \over \e$ term in this expansion is a bit long and is relegated to Appendix B. Using the result of this Appendix and noting that the prefactors in (\ref{xzd}) are just $S^{(0)} $ and $S^{(1)}$ the final result is
\be \label{mainclai}
{\cal M}_{n+1}(...,\{ \epsilon \lambda_s, \tilde\lambda_s\}) = \left({1 \over \e^3}S^{(0)} + {1 \over \e^2}S^{(1)}+ {1 \over \e}S^{(2)}\right){\cal M}_n+\co (\epsilon^0 ).
\ee
This completes the proof of our main claim in the form (\ref{mainclaim}).

\section{Identities for stripped amplitudes}

The main result of the previous section is a relation between the soft limit expansion of an unstripped $n+1$-particle amplitude and differential operators acting on an unstripped $n$-particle amplitude. In this section we translate this into relations between stripped amplitudes. Such relations are easily applied to various explicit formulae for amplitudes appearing in the literature. We give some examples in the next section.

Stripped amplitudes are in general defined off of the momentum-conserving locus.  Therefore stripped relations must specify a continuation of the amplitudes off of this locus. For any $m$-particle amplitude the basic relation is
\be
{\cal M}_m = M_m ~\delta^4(\sum_{a=1}^mk_{a}^\mu).
\ee
Here $m$ is arbitrary; below we use both $m=n+1$ and $m=n$. A standard way of solving for momentum conservation in the spinor-helicity formalism is by selecting any two particles, say $1$ and $2$, to write (see e.g. \cite{Witten:2003nn})
\be
\delta^4(\sum_{a=1}^mk_{a}^\mu) = \frac{1}{\langle 1, 2\rangle^2}\delta^2\left( \sum_{a=1}^m \frac{\langle 1, a\rangle}{\langle 1, 2\rangle}\tilde\lambda_a\right)\delta^2\left( \sum_{a=1}^m \frac{\langle 2, a\rangle}{\langle 2, 1\rangle} \tilde\lambda_a \right).
\ee
On the support of these delta functions one has
\be\label{uniq}
\tilde\lambda_1 = - \sum_{a=3}^m \frac{\langle 2, a\rangle}{\langle 2, 1\rangle}\tilde\lambda_a, \quad\quad \tilde\lambda_2 = - \sum_{a=3}^m \frac{\langle 1, a\rangle}{\langle 1, 2\rangle} \tilde\lambda_a
\ee
which implies
\be \sum_{a=1}^m \lambda_a\tilde\lambda_a=0.\ee
A definition of $M_m$  for all $(\lambda_a, \tilde \lambda_a)$  can be obtained by equating it to its predetermined value on the momentum-conserving locus and extending it off the locus  by the simple condition
\be {\p \over \p \tilde\lambda_1 }M_m={\p \over \p \tilde\lambda_2}M_m=0.\ee
That is $M_m$ is taken to be a function of only the $2m-2$ spinors that can be varied consistently with the constraint of momentum conservation, and
is denoted by
\be
M_m^{(12)} = M_m^{(12)}(\{ \lambda_1,\lambda_2,\ldots ,\lambda_m\}, \{\tilde\lambda_3,\ldots ,\tilde\lambda_m\})
\ee
where the label $(12)$ indicates that the functions do not depend on the variables $\{ \tilde\lambda_1,\tilde\lambda_2\}$.

The relation among full amplitudes can then be expressed as a relation among elements in any such family of functions. The family of functions we are interested in is
$$\{ M_3^{(12)}, M_4^{(12)}, M_5^{(12)}, \ldots  \}$$
The explicit relation is then
\begin{eqnarray}\label{stClaim}
&& M_{n+1}^{(12)}(\{\pl_1,\pl_2 ,\ldots ,\pl_n, \epsilon\pl_s\},\{\tilde\pl_3,\tilde\pl_4, \ldots ,\tilde\pl_n,\tilde\pl_s\}) = \nonumber\\ && \left({1 \over \e^3}S^{(0)} + {1 \over \e^2}S^{(1)}+ {1 \over \e}S^{(2)}\right)M_n^{(12)}(\{\pl_1,\pl_2,\ldots ,\pl_n\},\{\tilde\pl_3,\tilde\pl_4, \ldots ,\tilde\pl_n\}) + \co (\epsilon^0 ).
\end{eqnarray}
It is this delta-function-free identity which will be verified in explicit examples in the next section.

\section{Low-point examples}

In the previous sections we presented formal proofs that the identity (\ref{nextsf}) and its stripped version (\ref{stClaim}) are valid for all graviton tree amplitudes.
It is both illuminating and fun to actually verify this in detail from the known explicit expressions for these amplitudes. In this section we verify (\ref{stClaim}) for all stripped amplitudes with six or fewer gravitons. This includes, in the last subsection,  the lowest order  next to maximally-helicity-violating (MHV) amplitude which arises for six gravitons.

\subsection{MHV amplitudes}

General MHV amplitudes of gravitons are known to be polynomials in the antiholomorphic spinors, $\tilde\lambda$. This means that the soft limit expansion terminates. In fact, $M^{\rm MHV}_{n+1}(\e )$, only contains $n-2$ terms. Since the soft theorems we discussed constrain the first three terms, amplitudes with $n+1 \leq 6$ are completely determined from (\ref{nextsf}).

The starting point is a three point amplitude, which is completely determined by the symmetries to be
\be
M_3(\{ k_1,-2\},\{k_2,-2\},\{k_3,+2\}) =\frac{\langle 1,2\rangle^6}{\langle 2,3\rangle^2\langle 3,1\rangle^2}.
\ee
There is a slightly more symmetric way of writing this amplitude which makes manifest the helicity dependence
\be
M_3(\{ k_1,-2\},\{k_2,-2\},\{k_3,+2\}) =\frac{\langle 1,2\rangle^8}{|1,2,3|^2}
\ee
with
\be
|1,2,3|\equiv \langle 1,2\rangle\langle 2,3\rangle\langle 3,1\rangle.
\ee

\subsubsection{Four gravitons}

Applying the soft expansion formula (\ref{mainclaim}) to $n+1=4$ with the soft particle chosen to be graviton $4$ one has,
\begin{eqnarray}
&& M_4^{(12)}(\{ \pl_1,\pl_2,\pl_3,\e \pl_4\},\{\tilde\pl_3,\tilde\pl_4\}) =\\
&& \left({1 \over \e^3}S^{(0)}+{1 \over \e^2}S^{(1)}+{1 \over \e}S^{(2)}\right)M_3^{(12)}(\{\pl_1,\pl_2,\pl_3\},\{\tilde\pl_3\}).
\end{eqnarray}
Note that since $M_3$ is purely holomorphic, it is annihilated by $S^{(1)}$ and $S^{(2)}$ and we trivially find
\be\label{claim4}
M_4^{(12)}(\e ) =\frac{1}{\epsilon^3}\frac{\langle 1,2\rangle^8}{|1,2,3|^2}\left(\sum_{a=1}^3\frac{[4,a]\langle x,a\rangle\langle y,a\rangle}{\langle 4,a\rangle\langle x,4\rangle\langle y,4\rangle}\right).
\ee
Very nicely, this formula is exact in $\e$ and therefore it can be evaluated at $\e=1$ to get the corresponding finite 4-point amplitude. The presentation of the four-particle amplitude in  (\ref{claim4}) is precisely the one obtained  by specializing Hodges' determinant formula \cite{Hodges:2011wm,Hodges:2012ym}, reviewed in appendix C equation (\ref{hodgesMHV}), to the four particle case.

\subsubsection{Five gravitons}

In this case the situation is more interesting as we will see $S^{(1)}$ in action. Here we take particle $5$ as the soft graviton.

In order to apply (\ref{mainclaim}) more easily to the four particle amplitude we choose to use the presentation found above in (\ref{claim4}) with reference spinors $x=1$ and $y=2$, i.e.
\be
M_4^{(12)} = \frac{\langle 1,2\rangle^8}{|1,2,3|^2}\frac{[4,3]\langle 1,3\rangle\langle 2,3\rangle}{\langle 4,3\rangle\langle 1,4\rangle\langle 2,4\rangle}.
\ee
Note that this representation has the virtue of not depending on either $\tilde\pl_1$ or $\tilde\pl_2$. It is also easy to see that $S^{(2)}$ annihilates the amplitude since it is only linear in $\tilde\pl$'s.

A short calculation gives
\be
S^{(1)}M_4^{(12)} = \frac{1}{|123|^2}\frac{[5,3][5,4]\langle 3,1\rangle\langle 3,2\rangle}{\langle 5,3\rangle\langle 5,4\rangle\langle 4,1\rangle\langle 4,2\rangle}.
\ee
We choose to use $S^{(0)}$ with reference spinors $x=1$ and $y=2$, i.e.
\be
S^{(0)} = \sum_{a=3}^4 \frac{[5,a]\langle 1,a\rangle\langle 2,a\rangle}{\langle 5,a\rangle\langle 1,5\rangle\langle 2,5\rangle}.
\ee
In order to compare with the $5$-point answer, we use Hodges' formula (\ref{hodgesMHV}) again. After performing the deformation and expanding in $\epsilon$ one has
\be
M_5^{(12)}(\epsilon ) = \frac{1}{\epsilon^3}S^{(0)}M_4^{(12)} + \frac{1}{\epsilon^2}\left(\frac{1}{|123|^2}\frac{[5,3][5,4]\langle 3,1\rangle\langle 3,2\rangle}{\langle 5,3\rangle\langle 5,4\rangle\langle 4,1\rangle\langle 4,2\rangle}\right).
\ee
where the $1/\e^2$ term precisely agrees $S^{(1)}M_4^{(12)}$. Once again, the formula for $M_5^{(12)}(\e )$ is exact in $\e$ and gives the full amplitude when evaluated at $\e =1$.

\subsubsection{Six gravitons}

 Here we take particle $6$ as the soft graviton.
Using Hodges' formula in the form (\ref{hodgesMHV}) for the five particle amplitude one finds
\be
S^{(2)}M_5^{(12)} = -S^{(0)}G_4G_5-G_5\frac{[4,6]^2}{\langle 4,6\rangle^2} - G_4\frac{[5,6]^2}{\langle 5,6\rangle^2}
\ee
where
\be
S^{(0)} & = & \sum_{a=3}^5 \frac{[6,a]\langle 1,a\rangle\langle 2,a\rangle}{\langle 6,a\rangle\langle 1,6\rangle\langle 2,6\rangle}, \\
G_4 &=& -\frac{[4,6]\langle 1,6\rangle\langle 2,6\rangle}{\langle 4,6\rangle\langle 1,4\rangle\langle 2,4\rangle} , \\
G_5 &=& -\frac{[5,6]\langle 1,6\rangle\langle 2,6\rangle}{\langle 5,6\rangle\langle 1,5\rangle\langle 2,5\rangle}.
\ee

Once again, applying the deformation to Hodges' formula for six gravitons (\ref{hodgesMHV}) in the Appendix one finds perfect agreement
\be
M_6^{(12)}(\epsilon ) = \frac{1}{\epsilon^3}S^{(0)}M_5^{(12)} + \frac{1}{\epsilon^2}S^{(1)}M_5^{(12)} + \frac{1}{\epsilon}S^{(2)}M_5^{(12)}.\ee
The formula for six MHV gravitons is also exact in $\epsilon$.
Starting at seven MHV gravitons we do expect to have $\co (\epsilon^0)$ corrections.  We also expect to find such corrections for the six point next-to-MHV (NMHV) amplitude which we discuss next.

\subsection{Six-graviton NMHV }

This example is the most interesting one as it requires the use of every detail in (\ref{stClaim}). The starting point is the five point amplitude with helicities $\{ h_1=-2,h_2=-2,h_3=-2,h_4=+2,h_5=+2\}$. This is obtained from Hodges'  formula (\ref{hodgesMHV}) by conjugation
\be
M_5 = \frac{[4,5]^6}{[3,4]^2[5,3]^2}\left(\tilde\phi_{11}\tilde\phi_{22}-\tilde\phi_{12}^2\right)
\ee
with
\be
\tilde\phi_{aa} = -\sum_{b=1}^5\frac{\langle a,b\rangle[b,4][b,5]}{[a,b][a,4][a,5]}, \quad  \tilde\phi_{12} = \frac{\langle 1,2\rangle}{[1,2]}.
\ee
In order to apply (\ref{stClaim}), we construct $M^{(1,2)}_5$ by using momentum conservation to eliminate $\tilde\lambda_1$ and $\tilde\lambda_2$, more precisely,
\be
M^{(12)}_5(\{\lambda_1,\lambda_2, \ldots ,\lambda_5\},\{\tilde\lambda_3,\tilde\lambda_4,\tilde\lambda_5\}) =
M_5(\{\lambda_1,\tilde\lambda_1^*\},\{\lambda_2,\tilde\lambda_2^*\},\{ \lambda_3,\tilde\lambda_3\}, \ldots )
\ee
with
\be
\tilde\lambda_1^*= -\sum_{a=3}^5 \frac{\langle 2, a\rangle}{\langle 2, 1\rangle}\tilde\lambda_a,\quad \tilde\lambda_2^* = - \sum_{a=3}^5 \frac{\langle 1, a\rangle}{\langle 1, 2\rangle}\tilde\lambda_a .
\ee
It is now easy to apply (\ref{stClaim}) to obtain
\be
\left({1 \over \e^3}S^{(0)} + {1 \over \e^2}S^{(1)}+ {1 \over \e}S^{(2)}\right)M_5^{(12)}(\{\pl_1,\pl_2,\ldots ,\pl_5\},\{\tilde\pl_3,\tilde\pl_4, \tilde\pl_5\}).
\ee
In order to compare to the six-point amplitude we take the explicit form given in \cite{Cachazo:2005ca} as
\be
M_6 = D_1+{\overline D_1}^{\rm flip}+D_2+D_3+{\overline D_3}^{\rm flip}+D_6
\ee
with\footnote{Here we use the opportunity to point out a typographical error in the formula for $D_1$ given in \cite{Cachazo:2005ca} where the second term in the numerator must have a minus sign.}
\begin{eqnarray}
D_1 &=& \frac{\langle 23\rangle\langle 1|2+3|4]^7(\langle 1|2+3|4]\langle 5|3+4|2][51]-[12][45]\langle 51\rangle s_{234})}{\langle 15\rangle\langle 16\rangle[23][34]^2\langle 56\rangle s_{234}\langle 1|3+4|2]\langle 5|3+4|2]\langle 5|2+3|4]\langle 6|3+4|2]\langle 6|2+3|4]}\nonumber \\ && +(1\leftrightarrow 2).
\end{eqnarray}
Here $s_{234}=(k_2+k_3+k_4)^2$ and $\langle a|b+c|d] = \langle a,b\rangle[b,d]+ \langle a,c\rangle[c,d]$. In the formula for $M_6$, ${\overline D_1}^{\rm flip}$ means taking the formula for $D_1$ exchange $[~,~]\leftrightarrow \langle ~,~\rangle$ and relabel $i\to 7-i$.

The remaining pieces are given by
\begin{eqnarray}
D_2 &=& -\frac{\langle 13\rangle^7\langle 25\rangle[45]^7[16]}{\langle 16\rangle[24][25]\langle 36\rangle s_{245}\langle 1|2+5|4]\langle 6|2+5|4]\langle 3|1+6|5]\langle 3|1+6|2]}\nonumber \\
&& + (1\leftrightarrow 2)+(5\leftrightarrow 6)+(1\leftrightarrow 2,5\leftrightarrow 6),
\end{eqnarray}
\begin{eqnarray}
D_3 &=& \frac{\langle 13\rangle^8[14][56]^7(\langle 23\rangle\langle 56\rangle[62]\langle 1|3+4|5]+\langle 35\rangle[56]\langle 62\rangle\langle 1|3+4|2])}{\langle 14\rangle[25][26]\langle 34\rangle^2s_{134}\langle 1|3+4|2]\langle 1|3+4|5]\langle 1|3+4|6]\langle 3|1+4|2]\langle 3|1+4|5]\langle 3|1+4|6]}\nonumber \\
&& (1\leftrightarrow 2),
\end{eqnarray}
and
\begin{eqnarray}
D_6 &=& \frac{\langle 12\rangle[56]\langle 3|1+2|4]^8}{[21][14][24]\langle 35\rangle\langle 36\rangle\langle 56\rangle s_{124}\langle 5|1+2|4]\langle 6|1+2|4]\langle 3|5+6|1]\langle 3|5+6|2]}.
\end{eqnarray}

The next step is to construct the function $M^{(12)}_6$ obtained from $M_6$ as
\be
M^{(12)}_6(\{\lambda_1,\lambda_2, \ldots ,\lambda_6\},\{\tilde\lambda_3,\tilde\lambda_4,\ldots, \tilde\lambda_6\}) =
M_6(\{\lambda_1,\tilde\lambda_1^*\},\{\lambda_2,\tilde\lambda_2^*\},\{ \lambda_3,\tilde\lambda_3\}, \ldots )
\ee
with
\be
\tilde\lambda_1^*= -\sum_{a=3}^6 \frac{\langle 2, a\rangle}{\langle 2, 1\rangle}\tilde\lambda_a,\quad \tilde\lambda_2^* = - \sum_{a=3}^6 \frac{\langle 1, a\rangle}{\langle 1, 2\rangle}\tilde\lambda_a .
\ee

Finally, we construct a rational function of $\epsilon$
\be
M_6^{(12)}(\{\lambda_1,\lambda_2, \ldots ,\epsilon\lambda_6\},\{\tilde\lambda_3,\tilde\lambda_4,\ldots, \tilde\lambda_6\})
\ee
and expand around $\epsilon =0$. In this case, we find that the series does not terminate and that the first three orders are exactly given by
\be
\left({1 \over \e^3}S^{(0)} + {1 \over \e^2}S^{(1)}+ {1 \over \e}S^{(2)}\right)M_5^{(12)}(\{\pl_1,\pl_2,\ldots ,\pl_5\},\{\tilde\pl_3,\tilde\pl_4, \tilde\pl_5\})
\ee
as expected.

\section*{Acknowledgements}

We are grateful to  N. Arkani-Hamed, J. Bourjaily, D. Kapec, V. Lysov, S. Pasterski, X. Yin and E. Yuan for useful conversations.
FC is supported by Perimeter Institute for Theoretical Physics. Research at Perimeter Institute is supported by the Government of Canada through Industry Canada and by the Province of Ontario through the Ministry of Research \& Innovation.  AS is supported in part by NSF grant 1205550.

\appendix

\section{Absence of poles in $\cal R$}

A crucial step in the construction of section 2 is the finiteness of $\cal R$ as $\epsilon\to 0$. In this appendix we prove this statement. Recall that the terms contributing to $\cal R$ come from poles of $M_{n+1}(z)/z$ located at solutions of
\be\label{condx}
(k_s(z)+k_{a_1}+\cdots k_{a_m})^2=0
\ee
with $m>1$. Let us denote as $Q_m$ the vector $k_{a_1}+\cdots +k_{a_m}$. The solution to equation (\ref{condx}) is
\be
z^* = -\frac{(k_s+Q_m)^2}{2p\cdot Q_m} \quad {\rm where} \quad p_{\alpha,\dot\alpha} = \lambda_{n,\alpha}\tilde\lambda_{s,\dot\alpha}.
\ee
Here we used that $p\cdot k_s = 0$.

The terms contributing to ${\cal R}$ are then of the form
\be
\sum_{h_I=\pm 2}M_L(s(z^*),a_1,\ldots ,a_m,\{k_I,h_I\})\frac{1}{(k_s+Q_m)^2}M_R(\{-k_I,-h_I\},n(z^*),\overline{a_1,\ldots , a_m}).
\ee
Consider the limit $\lambda_s\to 0$. In that case
\be
z^* = -\frac{Q_m^2}{2p\cdot Q_m}, \quad k_I = Q_m -\frac{Q_m^2}{2p\cdot Q_m} p.
\ee
Note that $p$ remains finite since $\tilde\lambda_s$ is held fixed. The momentum of the internal particle, i.e. $k_I$,  is the projection of $Q_m$ along the null direction defined by $p$ and hence is itself null.

Let us consider the left amplitude in the limit $\lambda_s\to 0$,
\be
M_L \to M_L(z^*p,a_1,\ldots ,a_m,\{k_I,h_I\}) .
\ee
This amplitude is evaluated at generic momenta away from any soft, collinear or multiparticle factorization singularities and hence it is finite. Note that here $Q_m^2\neq 0$. This is precisely what is not true for the terms that do not belong to $\cal R$ and were studied in detail in section 2.3.1.

In the holomorphic soft limit the right amplitude becomes
\be
M_R \to M_R(\{-k_I,-h_I\},k_n(z^*),\overline{a_1,\ldots , a_m})
\ee
where
\be
k_n(z^*) = k_n-z^*p.
\ee
Clearly, if the number of elements in $\overline{a_1,\ldots , a_m}$ is greater than two then the amplitude is again evaluated away from any singularities and hence finite. The only possibly dangerous case is when $\overline{a_1,\ldots , a_m}$ consists of a single element. Let us assume that such a particle is particle $b$. In this case we have
\be
M_R(\{k_I,-h_I\},\{ k_n(z^*), h_n \},\{ k_b, h_b\}).
\ee
We can evaluate all kinematic invariants explicitly by using momentum conservation to write $Q =-k_b-k_s-k_n$. Moreover, since $\lambda_s\to 0$ we have $Q=-k_b-k_n$. A short computation also reveals that
\be
k_I = \left(-\lambda_b-\frac{[s,n]}{[s,b]}\lambda_n\right)\tilde\lambda_b,
\ee
\be
k_n(z^*) = \frac{[s,n]}{[s,b]}\lambda_n\tilde\lambda_b.
\ee
Note that all three momenta entering in $M_R$ have $\tilde\lambda_b$ by antiholomorphic spinor. This means that when the helicities are such that the amplitude has two pluses and one negative helicity then it vanishes. When the amplitude has two negative and one plus then one gets something finite. In either case the answer is non-singular and hence $\cal R$ is finite in the soft holomorphic limit $\lambda_s\to 0$ as claimed.

\section{Second-order Taylor expansion}

In this appendix we complete the argument at the end of section 2.3.2 by expanding the near-soft amplitude to second order in $\e$.

Consider the distribution
\be
G(\epsilon ) ={\cal M}_n(\{\lambda_1,\tilde\lambda_1\},\ldots ,\{\lambda_a,\tilde\lambda_a+\epsilon\frac{\langle n,s \rangle}{\langle n,a \rangle}\tilde\lambda_s\},\ldots ,\{\lambda_n,\tilde\lambda_n+\epsilon\frac{\langle a,s \rangle}{\langle a,n \rangle}\tilde\lambda_s\} )
\ee
and expand it as
\be
G(\epsilon) = G(0)+G'(0)\epsilon +\frac{1}{2}G''(0)\epsilon^2 + \co (\epsilon^3)
\ee
The derivative with respect to $\epsilon$ becomes by applying the chain rule
\be
\frac{\partial}{\partial\epsilon} = \frac{\langle n,s \rangle}{\langle n,a \rangle}\tilde\lambda^{\dot\alpha}_s\frac{\partial}{\partial\tilde\lambda_a^{\dot\alpha}}+\frac{\langle a,s \rangle}{\langle a,n \rangle}\tilde\lambda^{\dot\alpha}_s\frac{\partial}{\partial\tilde\lambda_n^{\dot\alpha}}.
\ee
This implies that
\be
G''(0) = \left(\frac{\langle n,s \rangle^2}{\langle n,a \rangle^2}\tilde\lambda^{\dot\alpha}_s\tilde\lambda^{\dot\beta}_s\frac{\partial^2}{\partial\tilde\lambda_a^{\dot\alpha}
\partial\tilde\lambda_a^{\dot\beta}} - 2\frac{\langle a,s \rangle\langle n,s \rangle}{\langle a,n \rangle^2}\tilde\lambda^{\dot\alpha}_s\tilde\lambda^{\dot\beta}_s\frac{\partial^2}{\partial\tilde\lambda_n^{\dot\alpha}
\partial\tilde\lambda_a^{\dot\beta}} + \frac{\langle a,s \rangle^2}{\langle a,n \rangle^2}\tilde\lambda^{\dot\alpha}_s\tilde\lambda^{\dot\beta}_s\frac{\partial^2}{\partial\tilde\lambda_n^{\dot\alpha}
\partial\tilde\lambda_n^{\dot\beta}}\right){\cal M}_n.
\ee
The next step is to plug this on the right hand side of the $\epsilon$ deformed version of (\ref{repre}),namely %
\be\label{reprex}
{\cal M}_{n+1}(\epsilon )\!\! =\!\! \frac{1}{\epsilon^3}\sum_{a=1}^{n-1} \frac{[s,a]\langle n,a \rangle^2}{\langle s,a \rangle\langle n,s \rangle^2} {\cal M}_{n}(\{\lambda_1,\tilde\lambda_1\},\ldots ,\{\lambda_a,\tilde\lambda_a\! +\!\epsilon\frac{\langle n,s \rangle}{\langle n,a \rangle}\tilde\lambda_s\},\ldots ,\{\lambda_n,\tilde\lambda_n\! +\! \epsilon\frac{\langle a,s \rangle}{\langle a,n \rangle}\tilde\lambda_s\} ).\nonumber
\ee
The sub-subleading term is then
\begin{eqnarray}\label{nncc}
&& \frac{1}{2\epsilon} \sum_{a=1}^{n-1} \frac{[s,a]}{\langle s,a \rangle}\tilde\lambda^{\dot\alpha}_s\tilde\lambda^{\dot\beta}_s\frac{\partial^2}{\partial\tilde\lambda_a^{\dot\alpha}
\partial\tilde\lambda_a^{\dot\beta}}{\cal M}_n - \frac{1}{\epsilon} \sum_{a=1}^{n-1} \frac{[s,a]}{\langle s,n \rangle}\tilde\lambda^{\dot\alpha}_s\tilde\lambda^{\dot\beta}_s\frac{\partial^2}{\partial\tilde\lambda_a^{\dot\alpha}
\partial\tilde\lambda_n^{\dot\beta}}{\cal M}_n \nonumber\\
&& +\frac{1}{2\epsilon} \sum_{a=1}^{n-1} \frac{[s,a]\langle a,s \rangle}{\langle n,s \rangle^2}\tilde\lambda^{\dot\alpha}_s\tilde\lambda^{\dot\beta}_s\frac{\partial^2}{\partial\tilde\lambda_n^{\dot\alpha}
\partial\tilde\lambda_n^{\dot\beta}}{\cal M}_n.
\end{eqnarray}
The first sum is almost what is needed to prove our claim, i.e.,
\be
\frac{1}{2\epsilon} \sum_{a=1}^{n-1} \frac{[s,a]}{\langle s,a \rangle}\tilde\lambda^{\dot\alpha}_s\tilde\lambda^{\dot\beta}_s\frac{\partial^2}{\partial\tilde\lambda_a^{\dot\alpha}
\partial\tilde\lambda_a^{\dot\beta}}{\cal M}_n = S^{(2)}{\cal M}_n - \frac{1}{2\epsilon} \frac{[s,n]}{\langle s,n \rangle}\tilde\lambda^{\dot\alpha}_s\tilde\lambda^{\dot\beta}_s\frac{\partial^2}{\partial\tilde\lambda_n^{\dot\alpha}
\partial\tilde\lambda_n^{\dot\beta}}{\cal M}_n.
\ee
Let us simplify the last sum in (\ref{nncc}) using
\be
\sum_{a=1}^{n-1}[s,a]\langle a,s \rangle = -2\sum_{a=1}^{n-1}q\cdot k_a = 2q\cdot k_n = -[s,n]\langle s,n \rangle
\ee
which gives
\be
\frac{1}{2\epsilon} \sum_{a=1}^{n-1} \frac{[s,a]\langle a,s \rangle}{\langle n,s \rangle^2}\tilde\lambda^{\dot\alpha}_s\tilde\lambda^{\dot\beta}_s\frac{\partial^2}{\partial\tilde\lambda_n^{\dot\alpha}
\partial\tilde\lambda_n^{\dot\beta}} \M_n= -\frac{1}{2\epsilon}\frac{[s,n]}{\langle s,n \rangle}\tilde\lambda^{\dot\alpha}_s\tilde\lambda^{\dot\beta}_s\frac{\partial^2}{\partial\tilde\lambda_n^{\dot\alpha}
\partial\tilde\lambda_n^{\dot\beta}}\M_n.
\ee
The second sum in (\ref{nncc}) can be simplified by writing it as
\be
- \frac{1}{\epsilon} \sum_{a=1}^{n-1} \frac{[s,a]}{\langle s,n \rangle}\tilde\lambda^{\dot\alpha}_s\tilde\lambda^{\dot\beta}_s\frac{\partial^2}{\partial\tilde\lambda_a^{\dot\alpha}
\partial\tilde\lambda_n^{\dot\beta}}{\cal M}_n  = - \frac{1}{\epsilon} \frac{1}{\langle s,n \rangle}\tilde\lambda^{\dot\beta}_s\frac{\partial}{
\partial\tilde\lambda_n^{\dot\beta}}\left(\tilde\lambda^{\dot\alpha}_s\tilde\lambda^{\dot\gamma}_s\sum_{a=1}^{n-1} \tilde\lambda_{a,\dot\gamma}\frac{\partial}{\partial\tilde\lambda_a^{\dot\alpha}
}\right){\cal M}_n.
\ee
Using global angular momentum conservation one has
\be
\tilde\lambda^{\dot\alpha}_s\tilde\lambda^{\dot\gamma}_s\sum_{a=1}^{n-1} \tilde\lambda_{a,\dot\gamma}\frac{\partial}{\partial\tilde\lambda_a^{\dot\alpha}
} = \frac{1}{2}\tilde\lambda^{\dot\alpha}_s\tilde\lambda^{\dot\gamma}_s\sum_{a=1}^{n-1}{\tilde J}^{(a)}_{\dot\alpha\dot\gamma} = -\frac{1}{2}\tilde\lambda^{\dot\alpha}_s\tilde\lambda^{\dot\gamma}_s{\tilde J}^{(n)}_{\dot\alpha\dot\gamma} = -[s,n]\tilde\lambda^{\dot\alpha}_s\frac{\partial}{\partial\tilde\lambda_n^{\dot\alpha}
},
\ee
which gives
\be
- \frac{1}{\epsilon} \sum_{a=1}^{n-1} \frac{[s,a]}{\langle s,n \rangle}\tilde\lambda^{\dot\alpha}_s\tilde\lambda^{\dot\beta}_s\frac{\partial^2}{\partial\tilde\lambda_a^{\dot\alpha}
\partial\tilde\lambda_n^{\dot\beta}}{\cal M}_n  = \frac{1}{\epsilon}\frac{[s,n]}{\langle s,n \rangle}\tilde\lambda^{\dot\alpha}_s\tilde\lambda^{\dot\beta}_s\frac{\partial^2}{\partial\tilde\lambda_n^{\dot\alpha}
\partial\tilde\lambda_n^{\dot\beta}}{\cal M}_n.
\ee
Combining all the contributions the final form of the sub-subleading term is
\be
\frac{1}{2\epsilon} \sum_{a=1}^{n} \frac{[s,a]}{\langle s,a \rangle}\tilde\lambda^{\dot\alpha}_s\tilde\lambda^{\dot\beta}_s\frac{\partial^2}{\partial\tilde\lambda_a^{\dot\alpha}
\partial\tilde\lambda_a^{\dot\beta}}{\cal M}_n = S^{(2)}{\cal M}_n
\ee
as desired.

\section{Hodges' MHV formula}

In two beautiful papers \cite{Hodges:2011wm,Hodges:2012ym}, Hodges constructed a formula for all MHV tree-level amplitudes which is very compact and makes Weinberg's soft limits manifest. It is therefore not surprising that this is our preferred choice of presentation for MHV amplitudes.

A stripped  $n$-particle MHV amplitude is given in terms of the determinant of an $n\times n$ matrix $\Phi$ with entries

\be\label{hodges}
\Phi_{a b}=\begin{cases} \displaystyle \frac{[a, b]}{\langle a,b\rangle}, \quad a\neq b,\\
\displaystyle -\sum_{c\neq a}\Phi_{a c}\frac{\langle x,c\rangle\langle y,c\rangle}{\langle x,a\rangle\langle y,a\rangle},\quad a=b,\end{cases}
\ee
where $x$ and $y$ represent two reference spinors.

Hodges showed that this matrix has corank 3, i.e., the dimension of the null space is three. He did so by finding an explicit basis for the null space. Using the basis Hodges showed that a permutation invariant formula of the determinant of any $(n-3)\times (n-3)$ minor of $\Phi$ is given by
\be
{\rm det}'\Phi \equiv \frac{\det \Phi^{ijk}_{pqr}}{|ijk||pqr|}
\ee
where $\Phi^{ijk}_{mpq}$ is the $(n-3)\times (n-3)$ matrix obtained from $\Phi$ by removing rows $i,j,k$ and columns $p,q,r$. The factors in the denominator are defined as
\be
|ijk| \equiv \langle i,j\rangle\langle j,k\rangle\langle k,i\rangle
\ee
and similarly for $|pqr|$.

Finally, the formula for an $n$-particle MHV amplitude with negative helicity gravitons $1$ and $2$ and the rest of positive helicity is given by
\be
M_n = \langle 1,2\rangle^8 {\rm det}'\Phi.
\ee
In the examples section we use a particularly convenient choice of columns and rows as well as a convenient gauge. The choice of rows and columns in $\{i,j,k\} = \{ p,q,r\} =\{ 1,2,3\}$ while the choice of gauge is $x=1$ and $y=2$. In this form
\be\label{hodgesMHV}
M_n =\frac{\langle 1,2\rangle^8}{|1,2,3|^2}\det \Phi_{123}^{123}.
\ee

\end{document}